# AN EFFICIENT IMPLEMENTATION OF THE OPTIMAL JET DEFINITION

**Contribution to the XIII Int. Workshop on Quantum Field Theory and High Energy Physics (QFTHEP'99, 27 May – 2 June 1999, Moscow)**

D. Yu. Grigoriev[a] and F. V. Tkachov[b]

*Institute for Nuclear Research of Russian Academy of Sciences*
*60th October Ave. 7a, Moscow 117312, Russia*

[a] dgr@ms2.inr.ac.ru,  [b] ftkachov@ms2.inr.ac.ru

We describe an efficient implementation of the optimal jet definition derived in hep-ph/9901444 (revision of January, 2000) and implemented in Fortran 77. The algorithm handles both c.m.s. and hadron collision kinematics.

## Introduction                                                                   1

Jet finding algorithms are a central data-processing tool in high-energy physics. However, the definition of jets suffered from misinterpreted ambiguities, which resulted in several existing jet finding schemes (cone, recombination, …), each in several variations (for a review see e.g. [1]). This resulted in practical difficulties since each experiment tends to use its own jet definition, making it difficult to compare physical results.

A systematic theoretical analysis of jet-related measurements from first principles was performed in [2], [3] where the central importance of the following two general requirements was pointed out:

(a) stability of data processing algorithms with respect to small effects such as errors in experimental data, etc.;

(b) compatibility with quantum field theory (the connection of generalized shape observables with the fundamental energy-momentum tensor was established in [4]).

The theory of [2], [3] criticized the conventional view on jet-finding algorithms as inversion of hadronization and, instead, offered to regard jets as a tool of approximate description of hadronic events — in full compliance with the first principles of physical measurements [2], [3]. This opened way for deriving a jet finding criterion entirely from first principles [5]. The criterion is derived from studying how physical information (represented by the mentioned fundamental shape observables) is distorted in the transition from particles to jets, and requiring that such a distortion is minimal.

The criterion of [5] can be regarded as a cone algorithm rewritten in terms of thrust-like shape observables, which makes it potentially very attractive.

In the practical aspect, however, the algorithmic usefulness of the definition [5] has remained somewhat problematic because it involved a minimization of a function within a bounded domain in the space of $N_\text{jets} \times N_\text{particles}$ dimensions with an additional restriction. With $N_\text{particles} = 100 \div 400$ (the numbers typical for LHC) and $N_\text{jets} = 1 \div 6$, one deals with a huge dimensionality (up to 2000 and more), which makes the resulting optimization problem non-trivial.

Fortunately, the analytical regularity of the criterion of [5] means there is a lot of information about the problem — enough to allow a quite efficient implementation of the corresponding jet-finding algorithm, apparently on a par with conventional algorithms.

The purpose of the present Letter is to describe such an algorithm with emphasis on analytical, programming-language-independent aspects. The currently available version is written in Fortran 77 (adapted from a code developed in Component Pascal [7]), has been compiled on a number of platforms, and works for both c.m.s. (spherically symmetric) kinematics, and for hadron-hadron collisions (cylindrical kinematics). The algorithm reliably solves the minimization problem of [5] (i.e. finds the optimal configuration of jets) for a typical event in a small fraction of a second on a modest workstation.

The source together with a practical description of interfaces and examples is available on the Web. We aimed at designing a generic robust algorithm and well-structured code with a sufficient number of interface hooks to allow one to perform further modifications in case of need. Here we just mention that the central subroutine is `Q_minimize`; it descends into a (local) minimum from a given initial configuration, and it can be used in different ways, depending on a specific application, such as finding all local minima, etc.; cf. the discussion of construction of jet-based observables in [5].

## The criterion 2

In the summary description given below we do not attempt to explain the meaning of, and motivations behind all the details of the definition; the reader is referred to [5].

The event **P** is represented as a collection of its "particles",

$$\mathbf{P} = \{E_a, \hat{\boldsymbol{p}}_a\}_a,\qquad 2.1$$

each particle described by its energy $E_a$ and direction $\hat{\boldsymbol{p}}_a$, the interpretation of which two quantities depends on the kinematics:

For spherical ($e^+e^-$) kinematics, $E_a$ is the energy of the particle, and $\hat{\boldsymbol{p}}_a$ is the corresponding unit 3-vector (which can be represented e.g. by the two standard angles $\varphi_a, \theta_a$).

For cylindrical (hadron collisions) kinematics, $E_a$ is the transverse energy (also denoted as $E_{\perp a} = E_a^{\text{true}} \sin\theta_a$), whereas $\hat{\boldsymbol{p}}_a$ is represented by the (pseudo) rapidity $\eta_a = \ln\cot(\theta_a/2)$, $-\infty < \eta_a < +\infty$ (the polar angle $\theta_a$ is measured from the beam axis), and the transverse direction $\hat{\boldsymbol{p}}_{\perp a}$ (unit 2-vector which is normal to the beam axis; it can be represented by $\varphi_a$).

In either case, the energies $E_a$ are normalized so that

$$\sum_a E_a = 1.\qquad 2.2$$

With each pair $E_a, \hat{\boldsymbol{p}}_a$, one associates a unique light-like Lorentz 4-vector $p_a, p_a^2 = 0$.

A configuration of jets **Q** is described in a similar fashion as a collection of jets' energies and directions:

$$\mathbf{Q} = \{\mathcal{E}_j, \hat{\boldsymbol{q}}_j\}_{j=1}^{N_{\text{jets}}}.\qquad 2.3$$

In the conventional algorithms, particles and jets are linked by indicating which particle belongs to which jet. The scheme of [5] allows more flexibility: one introduces the so-called *recombination matrix* $z_{aj}$ which describes the fraction of $a$-th particle which went into formation of the $j$-th jet. $z_{aj}$ can be any real number $0 \le z_{aj} \le 1$ such that

$$\sum_{j=1}^{N_{\text{jets}}} z_{aj} \le 1 \quad \text{for any } a.\qquad 2.4$$

The inequality corresponds to the fact that part of the particle's energy is allowed to not participate in jet formation. So it is convenient to introduce the quantity

$$\bar{z}_a \stackrel{\text{def}}{=} 1 - \sum_j z_{aj} \quad \text{for any } a.\qquad 2.5$$

Note that the conventional scheme corresponds to restricting $z_{aj}$ to the value 1 if the $a$-th particle belongs to the $j$-th jet, and to 0 in the opposite case.

The recombination matrix $z_{aj}$ is the fundamental unknown in our scheme. In particular, the quantities interpreted as jets' physical momenta $p_j$ are expressed as follows:

$$p_j = \sum_a z_{aj} p_a.\qquad 2.6$$

For each jet one also defines a light-like Lorentz 4-vector $\tilde{q}_j$, $\tilde{q}_j^2 = 0$ called the *jet's 4-direction*, as follows:

Spherical kinematics:

$$\tilde{q}_j = (1, \hat{\boldsymbol{q}}_j),\qquad 2.7$$

where $\hat{\boldsymbol{q}}_j = \boldsymbol{p}_j / |\boldsymbol{p}_j|$ is a unit 3-vector and $\boldsymbol{p}_j$ is the space-like component of $p_j$.

Cylindrical kinematics:

$$\tilde{q}_j = (\cosh\eta_j, \sinh\eta_j, \hat{\boldsymbol{q}}_j^\perp),\qquad 2.8$$

where $\hat{\boldsymbol{q}}_j^\perp = \boldsymbol{p}_j^\perp / |\boldsymbol{p}_j^\perp|$ is a unit 2-vector with $\boldsymbol{p}_j^\perp$ being the transverse component of $p_j$, and $\eta_j$ determined from the relation

$$\mathcal{E}_j \eta_j = \sum_a z_{aj} E_a \eta_a.\qquad 2.9$$

Further, one defines two functions:

$$\mathrm{Y}[\mathbf{P},\mathbf{Q}] = 2\sum_j p_j \cdot \tilde{q}_j, \quad \mathrm{E}_{\text{soft}}[\mathbf{P},\mathbf{Q}] = \sum_a \bar{z}_a E_a.\qquad 2.10$$

The latter is interpreted as the "soft energy" which does not take part in jet formation.

The criterion is as follows. One chooses $R > 0$ and defines

$$\Omega_R[\mathbf{P},\mathbf{Q}] = R^{-2}\mathrm{Y}[\mathbf{P},\mathbf{Q}] + \mathrm{E}_{\text{soft}}[\mathbf{P},\mathbf{Q}].\qquad 2.11$$

Then one chooses a (small; say, 0.01) number $\omega_{\text{cut}} > 0$ and finds $z_{aj}$ which minimizes $\Omega$ and satisfies the restriction

$$\Omega_R[\mathbf{P},\mathbf{Q}] < \omega_{\text{cut}}\qquad 2.12$$

with a minimal number of jets.

It turns out that this jet finding criterion is similar to conventional cone algorithms although the expression 2.11 is a shape observable that generalizes the thrust to any number of thrust (semi-)axes (see [5] for a detailed discussion). Correspondingly, the parameter $R$ is similar to the cone radius of the conventional cone algorithm (the standard value 0.7 often adopted in the conventional algorithms roughly corresponds to $R = 1$ in our case). However, jet shapes in our case are determined dynamically taking into account the global shape of energy flow of the event.

The physical meaning of $R$ is the maximal jet radius as measured by infinitesimally soft particles (i.e. such a particle is relegated to soft energy if it is farther than $R$ from any jet's axis).

The parameter $\omega_{\text{cut}}$ is analogous to the jet resolution parameter of conventional recombination algorithms — and, simultaneously, to the so-called f-cuts in conventional algorithms [8] because Eq. 2.12 imposes an upper bound on soft energy.

See [5] for a detailed discussion of all this.

## The algorithm 3

It is convenient to treat the "soft energy" formally as a special 0-th jet and denote $z_{0j} = \bar{z}_a$. Then

$$\sum_{j=0}^{N_{\text{jets}}} z_{aj} = 1 \quad \text{for any } a.$$ 3.1

The $N_{\text{jets}}$-dimensional region described by the restriction 3.1 is the standard $N_{\text{jets}}$-dimensional simplex. Thus the configuration $z_{aj}$ one has to find is a point in a $N_{\text{jets}} \times N_{\text{part}}$-dimensional region which is a direct product of $N_{\text{part}}$ $N_{\text{jets}}$-dimensional standard simplices.

The algorithm we found to work well is a hybrid of the gradient method and a coordinate-by-coordinate optimization as well as a heuristic based on the experimental finding (a posteriori supported by some theoretical arguments) that the minimum tends to be located on configurations with the matrix elements $z_{aj}$ taking the values of either zero or one, which corresponds to vertices of the simplices 3.1. The algorithm consists in iteratively performing minimization with respect to $z_{aj}$ within the simplex 3.1 for each particle $a$.

### Gradient minimization within the standard simplex 3.2

For simplicity denote $n = N_{\text{jets}}$ and let $z_j \equiv z_{aj}$. The corresponding $n$-dimensional vector is denoted as $z$. It is convenient to work in terms of finding a maximum rather than minimum. So we are going to find a maximum of the function $F(z) \equiv -\Omega_R(z)$ for $z$ within a domain $D$ described by simple linear inequalities ("the standard $n$-dimensional simplex"):

$$z_j \geq 0, \quad \sum_j z_j \leq 1.$$ 3.3

Sums over $j$ such as in 3.3 run over $j = 1, \ldots, n$ unless explicitly restricted.

The simplest algorithm of maximum search is to start from a candidate point $z$ and to find the next candidate point in the form

$$z \to z + \tau d$$ 3.4

where $\tau > 0$ is a number and $d$ describes a direction in which $F$ increases. We are not interested here in $\tau$ (see however Sec. 4) and focus on finding $d$ up to an overall scalar factor from the local properties of $F$ (its first derivatives at $z$) taking into account that $z + \tau d$ must remain within the boundaries of the domain $D$.

First suppose $z$ is an internal point of $D$. The function is locally represented as

$$F(z + \tau d) = F(z) + \tau (f \cdot d) + O(d^2)$$ 3.5

where $f_j = \partial F(z)/\partial z_j$ and

$$(f \cdot d) = \sum_j f_j d_j.$$ 3.6

A natural desire is to find the direction $d$ in which the linear function $(f \cdot d)$ increases fastest. However, in order to quantify such a desire, one has to define the notion of distance along each direction, and this involves an arbitrariness. For example, one could use euclidean distance and then the direction of fastest increase corresponds to a uniquely defined point of the unit sphere — but the choice of the euclidean metrics itself is not unique (cf. the figure). In general, any vector satisfying $(f \cdot d) > 0$ can be made the direction of fastest increase for some euclidean metrics. The only general heuristic is to choose $d$ in some simple way which is natural in the context of a specific problem. If the only information available is Eq. 3.5 then the usual choice is

$$d = f.$$ 3.7

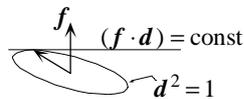

We adopt this choice for internal points of $D$, whereas the mentioned arbitrariness is made use of in determining $d$ in the case when $z$ is on the boundary of $D$.

Next suppose $z$ is a point of the boundary of $D$ for which

$$z_j = 0, \quad j \in B^\circ; \quad z_j > 0 \quad j \notin B^\circ; \quad \sum_j z_j < 1,$$ 3.8

where $B^\circ$ is a subset of $\{1, \ldots, n\}$. Then $d$ must obey the following restrictions:

$$d_j \geq 0, \quad j \in B^\circ.$$ 3.9

A simple choice for such $d$ is this:

```
IF j ∈ B° THEN d_j := MAX(0, f_j) ELSE d_j := f_j END.
```
3.10

Consider the case when $z$ sits on the front face of the simplex 3.3, i.e.

$$z_j > 0, \quad \sum_j z_j = 1.$$ 3.11

Then instead of 3.9 one has

$$\sum_j d_j \leq 0.$$ 3.12

Choose $1 \leq J \leq n$ and change coordinates by replacing $d_J$ with the new independent coordinate $d_0$:

$$d_0 = -\sum_j d_j, \quad d_J = -\sum_{j \neq J} d_j - d_0.$$ 3.13

Note the useful symmetry between all the components $d_j$, $j = 0, \ldots, n$ which is best seen from the relation

$$\sum_{j=0}^n d_j = 0.$$ 3.14

In terms of 3.13, the restriction 3.12 takes the standard form $d_0 \geq 0$. Re-express $(f \cdot d)$ in terms of the new independent coordinates $(d_{j \neq J}, d_0)$:

$$(f \cdot d) = \sum_j f_j d_j = \sum_{j \neq J} \bar{f}_j d_j + \bar{f}_0 d_0,$$ 3.15

where

$$\bar{f}_{j \neq J} = f_j - f_J, \quad \bar{f}_0 = -f_J.$$ 3.16

In terms of the coordinates $(d_{j \neq J}, d_0)$, a valid direction of increase for $(f \cdot d)$ is given by the following analog of 3.10:

```
IF   j = 0    THEN d_j ≡ d_0 := MAX(0, \bar{f}_0)
ELSIF j ≠ J   THEN d_j := \bar{f}_j END.
```
3.17

Lastly, $d_J$ is found from 3.13.

We can now consider the points of the boundary of $D$ which satisfy the most general set of restrictions:

$$z_j = 0, \quad j \in B^\circ; \quad z_j > 0, \quad j \notin B^\circ; \quad \sum_j z_j < 1. \quad 3.18$$

This can be represented in a symmetric fashion by introducing

$$z_0 = 1 - \sum_j z_j. \quad 3.19$$

Then let $B = B^\circ \cup \{0\} \subset \{0, \ldots, n\}$, and Eq. 3.18 becomes

$$z_j = 0, \quad j \in B; \quad z_j > 0, \quad j \notin B. \quad 3.20$$

Note that $B \neq \{0, \ldots, n\}$ (no point on the boundary of the simplex can belong to all its faces simultaneously). Therefore, one can always choose $J \notin B$ and consider $d_j$, $j \neq J$ as independent variables (here $j$ can take the value 0). Then from 3.15 one deduces the following prescription for choosing $\boldsymbol{d}$:

$$\texttt{IF } j \in B \texttt{ THEN } d_j := \texttt{MAX}(0, \bar{f}_j)$$
$$\texttt{ELSIF } j \neq J \texttt{ THEN } d_j := \bar{f}_j \texttt{ END}. \quad 3.21$$

Lastly, $d_J$ is found from 3.13.

The above choices of $\boldsymbol{d}$ are not unique as seen from the arbitrariness in the choice of $J$.

### Formulas for derivatives 3.22

Here are the relevant formulas for derivatives of $Y$ with respect to $z_{aj}$:

$$\frac{\partial}{\partial z_{aj}} Y(\{z\}) = 2\left(\frac{\partial}{\partial z_{aj}} p_j\right)\tilde{q}_j + 2p_j\left(\frac{\partial}{\partial z_{aj}}\tilde{q}_j\right)$$
$$= 2p_a\tilde{q}_j + 2p_j\left(\frac{\partial}{\partial z_{aj}}\tilde{q}_j\right). \quad 3.23$$

Only one term in the sum over $j$ survives because the terms which correspond to $j' \neq j$ depend only on $z_{aj'}$.

To evaluate the second term one has to use specific expressions for each of the two standard kinematics. For spherical kinematics we obtain

$$p_j\left(\frac{\partial}{\partial z_{aj}}\tilde{q}_j\right) = -p_j\frac{\partial}{\partial z_{aj}}\hat{q}_j$$
$$= -|p_j|^{-1}\left[(p_j \cdot p_a) - (p_j \cdot \hat{q}_j)(\hat{q}_j \cdot p_a)\right] \equiv 0. \quad 3.24$$

For cylindrical kinematics we obtain:

$$p_j\left(\frac{\partial}{\partial z_{aj}}\tilde{q}_j\right) = p_j^0\left(\frac{\partial}{\partial z_{aj}}\tilde{q}_j^0\right) - p_j^z\left(\frac{\partial}{\partial z_{aj}}\tilde{q}_j^z\right) - \boldsymbol{p}_j^\perp\left(\frac{\partial}{\partial z_{aj}}\tilde{\boldsymbol{q}}_j^\perp\right)$$
$$= \left(\frac{\partial \eta_j}{\partial z_{aj}}\right)\left[p_j^0 \sinh\eta_j - p_j^z \cosh\eta_j\right], \quad 3.25$$

where

$$\frac{\partial \eta_j}{\partial z_{aj}} = E_{\perp j}^{-1} E_{\perp a}(\eta_a - \eta_j). \quad 3.26$$

The resulting formulas are sufficiently simple not to present calculational difficulties in either case.

## Implementation 4

We limit our discussion here to language-independent aspects. Specific interfaces and code examples are provided separately [6]. We only note that the design of algorithm (which required experimentation with data structures and interactive experimentation with the control parameters of the algorithm) was performed using the freely available RAD tool BlackBox Component Builder [7] based on an extremely simple, type-safe, object-oriented and efficient compiled language Component Pascal (of the distinguished Pascal/Modula-2/Oberon-2 pedigree [9]). The final algorithm turned out to be simple enough to allow a port to Fortran with some improvements resulting from experimentation with realistic test samples of events. It should be emphasized that the design of the algorithm would have been much harder without all the safety features and the stunning combination of power and simplicity of Component Pascal, and without the simplicity, high interactivity and GUI features of the BBCB.

Concerning our test samples of events, we used the total of 500 events generated using Jetset [10] for typical processes studied at CERN and FNAL. It was not our aim to arrive at any physical conclusions, and in fact the specific nature of events played practically no role because our algorithm is fairly generic, and its overall behavior is essentially insensitive to details of structure of events. The tests were performed only for numerical debugging, not any studies of physics. A final adjustment of some numerical parameters was made possible by a large-scale test run on a realistic event sample performed by Pablo Achard [11].

No comparison with conventional algorithms has been attempted yet (the situation has changed by 2001 — *FT*).

The minimization scheme described above is easily and straightforwardly implemented using only static data structures (easily mapped to Fortran arrays), among which the central are the 2-dimensional array $z_{aj}$ and the 1-dimensional array corresponding to the direction $\boldsymbol{d}$. The total data size is determined by the size of $z_{aj}$. The number of particles cannot exceed a few thousands, and the number of jets, a dozen or so. So the size of $z_{aj}$ is $O(1000) \times O(10) \times (8 \text{ Bytes}) \sim O(100K)$. If each subarray $z_{aj}$ for fixed $a$ is contained in a contiguous memory block then the internal loop (which corresponds to minimization with respect to one particle's parameters) always deals with $O(1K)$ of contiguous data, which ensures a very good localization of the algorithm and therefore a fast performance.

Concerning the ambiguity in the choice of $\boldsymbol{d}$ according to the formulas given in Sec. 3.2, we found it advantageous to perform maximization of the length of $\boldsymbol{d}$ (measured according to the simple norm $|\boldsymbol{d}| = \max_{j=0,\ldots n}|d_j|$ which is natural in the context of simplicial geometry) with respect to $J$ (which is a free parameter in the above formulas). Such an optimization involves a small amount of well-localized data and code involving only very simple operations, resulting in a fast execution, whereas the resulting overall speedup proved to be significant.

The choice of step length $\tau$ (cf. 3.4) is determined by the experimental finding that the minimum tends to be located at the boundary of the simplex. So we find $\tau$ from the requirement that the new candidate minimum $z + \tau\boldsymbol{d}$ for given $z$ and $d$ should be located at the boundary of the simplex, and if this results in an increase in the value of $\Omega$, then $\tau$ is iteratively divided by a constant factor ~3 until a minimum is found.

An important technical implementation detail (the so-called "snapping") concerns how one deals with the boundaries of the simplex: if some $z_{aj}$ is small enough (i.e. $z$ is close enough to a boundary of the simplex) then it is set to zero ("snapped" to the boundary). A similar snapping is used for the direction $d$. Such snappings are necessary because one needs to detect the situations when $z$ is at the boundary and the direction runs parallel to the boundary.

There are no difficulties with the termination condition: since the resulting minimum is located at the boundary of the region (we have not seen exceptions so far, and some analytical arguments seem to indicate that such exceptions can never occur), the minimum tends to "snap" to the boundary quite fast. Also, most particles find their jet pretty fast, and later iterations involve decreasingly smaller numbers of particles.

The fact that the trajectory of the search tends to travel along the boundary of the region makes the algorithm similar to the well-known simplex algorithm of combinatorial optimization but in our case the algorithm is sped up by reliance on explicit analytical formulas to determine the direction of the next candidate minimum.

Typically, the algorithm arrives at a (local) minimum from a randomly generated starting point in $O(100)$ iterations, in a fraction of a second on a Pentium II.

In general, for a given event the criterion may have more than one local minimum. This is discussed in detail in [5]. To find the global minimum it proved sufficient to repeat the search a few times starting from new randomly generated configurations $z_{aj}$. "A few times" depends on the character of one's events: 2 (or even 1) may be enough for most situations with hard jets, and 10 seemed to be sufficient for events corresponding to the LEP2 process $e^+e^- \to W^+W^- \to 4$ jets. The number of repeated searches anticorrelates with the fraction of events for which the algorithm fails to correctly identify global minimum. This number is therefore tied to the overall precision of the physical problem. The implementation also provides for an explicit specification of the initial configuration to allow e.g. output from a conventional algorithm to be used for that purpose.

Further optimizations are possible by way of adding more intellect/memory to the algorithm (e.g. giving priority in minimization to some particles, or using special heuristics to reduce the number of repeated searches in situations where several local minima may occur) but we felt that in view of a good speed of the minimum search the additional complexity is not warranted at this stage. So we limited the design to a generic algorithm while providing the modularity to allow one to build such improved algorithms in case of real need.

To summarize, the described implementation proves practical feasibility of the jet definition of [5], and the developed software allows easy modifications to accommodate further data processing options described in [5]. Such options, while potentially important in specific applications from physical viewpoint, are not expected to require major changes of the described minimization algorithm.

*Acknowledgments.* We are indebted to A.K.Skasyrskaya for generating test samples of events using Jetset and Pythia [10]. We thank P. Achard [11] for running useful large-scale tests and pointing out a misprint in this text. This work was supported in part by the Russian Foundation for Basic Research under grant 99-02-18365. FT thanks Oberon microsystems, Inc. [7] for providing their fabulous BlackBox Component Builder without which the finding of the described algorithm would have not been feasible.